\documentclass[12pt,oneside,english]{amsart}
\usepackage[T1]{fontenc}
\usepackage[latin9]{inputenc}
\usepackage{geometry}
\usepackage{amstext}
\usepackage{amsthm}
\usepackage{graphicx}
\usepackage{setspace}
\makeatletter
\numberwithin{equation}{section}
\numberwithin{figure}{section}
\theoremstyle{plain}
\newtheorem{thm}{\protect\theoremname}
\theoremstyle{plain}
\newtheorem{prop}[thm]{\protect\propositionname}
\theoremstyle{remark}
\newtheorem{rem}[thm]{\protect\remarkname}
\theoremstyle{definition}
\newtheorem{defn}[thm]{\protect\definitionname}

\usepackage{babel}
\usepackage{lineno}

\makeatother

\usepackage{babel}
\providecommand{\definitionname}{Definition}
\providecommand{\propositionname}{Proposition}
\providecommand{\remarkname}{Remark}
\providecommand{\theoremname}{Theorem}

\begin{document}

\title{True Epidemic Growth Construction Through Harmonic Analysis }
\maketitle
\begin{center}
STEVEN G. KRANTZ{\large{}}\footnote{Department of Mathematics, Washington University in St. Louis, Missouri,
USA. 

Email: sk@math.wustl.edu}{\large\par}
\par\end{center}

\begin{center}
PETER POLYAKOV{\large{}}\footnote{Department of Mathematics, University of Wyoming, Laramie, Wyoming,
USA. 

Email: polyakov@uwyo.edu}{\large\par}
\par\end{center}

\begin{center}
ARNI S.R. SRINIVASA RAO{\large{}}\footnote{Laboratory for Theory and Mathematical Modeling, Division of Infectious
Diseases and Division of Epidemiology, Medical College of Georgia
and Department of Mathematics, Augusta University, Augusta, Georgia,
USA. Email: arrao@augusta.edu}{\large\par}
\par\end{center}

\vspace{2cm}
\begin{abstract}
In this paper, we have proposed a two phase procedure (combining discrete
graphs and wavelets) for constructing a true epidemic growth. In the
first phase graph theory based approach was developed to update partial
data available and in the second phase we used this partial data to
generate a plausible complete data through wavelets. This procedure
although novel and implementable, still leave some questions unanswered. 
\end{abstract}



\vspace{1cm} 

\section{\textbf{Basis, Motivation and Introduction }}

In general, it is not easy to build the true epidemic growth curve
in a timely fashion for any newly emerging epidemic and the chances
of building the true growth picture worsens with poor disease reporting.
As we know, preparedness for an epidemic spread is a primary public
health concern for any health departments. Epidemic reporting of a
disease is a fundamental event in understanding two key parameters
in epidemiology, namely, epidemic diffusion within a population and
growth at a population level. Normally, for a real-time epidemic the
reporting of cases is rarely complete, especially if the epidemic
is new or symptoms are unknown or symptoms are yet to be discovered.
For viruses with shorter incubation periods without virus shedding
during the incubation period, any delay in reporting or lack of reporting
could lead to a severe epidemic due to absence of controlling measures.
For example, for ebola, the average incubation period is between 2
and 21 days, and an individual diagnosed with the ebola virus does
not spread the virus to others during this period. Suppose some of
these individuals with ebola are not diagnosed (and hence not reported
to the health care facilities); then, after 21 days, these individuals
will (unknowingly) spread ebola to others. There are other viruses
whose incubation period is small but they are contagious during this
period, for example influenza. Even for epidemics with established
symptoms, the reporting could be nowhere close to complete and the
impact of reporting on epidemic surveillance can then be theoretically
measured (Rao, 2012). 

In this study, we are trying to attempt a classical problem in epidemic
reporting within a novel framework within harmonic analysis principles.
This study develops methodologies of constructing complete data from
a partial data using wavelets.

\section{\textbf{Fundamental questions }}

We are raising here very fundamental questions in epidemic reporting.
For example, does an epidemic case reporting over time follow any
pattern? Or, in any particular situation, does an epidemic reporting
pattern have anything to do with the actual epidemic wave? Actual
epidemic or a true epidemic wave is the number of all cases (reported
and not reported) as a function of time. Is there any strong or weak
association between ``epidemic reporting patterns'' and ``epidemic
waves'' in general? Suppose we cannot generalize such an association
for every epidemic, then will there be any such association for a
particular epidemic?

In any epidemic, we can hardly observe the actual (true) epidemic
wave, and what we construct as a wave is mostly based on reporting
numbers of disease cases. The central questions in which we are interested
can be summarized as follows: 
\begin{enumerate}
\item[{\bf (i)}] How far the reporting of an epidemic is helping us in accurate prediction
of an epidemic, especially if it is an emerging epidemic? and how
far such an association is clarified (which it is not otherwise) using
methods of harmonic analysis? 
\item[{\bf (ii)}] It is seldom that the�cases generated in a population are completely
detected, so the question that remains unanswered during most of the
time in a newly emerging epidemic is: will there be any way to back-calculate
and reproduce these numbers lost in detection and, if so, to what
extent can we reconstruct accurately an epidemic growth (before control
measures are implemented)? 
\end{enumerate}
There are other related questions but first we want to see from the
lens of wavelet/harmonic/PDE analysis because we believe there could
be some useful light to be unearthed in this way.

Hence, it is always challenging to construct true epidemic waves because
population vaccination and control policies depend on understanding
the true ground level reality of disease cases.

\section{\textbf{Wavelets}}

In the past thirty-five years there has developed a new branch of
harmonic analysis called \textit{wavelet theory}\textemdash see (Meyer/Ryan,
1993), (Meyer, 1998), (Hernandez/Weiss, 1996), (Walker, 1997), (Strichartz,
1993), (Labate/Weiss/Wilson, 2013). Largely based on the ideas of
Yves Meyer, wavelet theory replaces the traditional Fourier basis
of sine functions and cosine functions with a more flexible and adaptable
basis of wavelets. The advantages of wavelets are these: 
\begin{enumerate}
\item[{\bf (a)}] The wavelet expansion of a function can be localized both in the
time and the space variable; 
\item[{\bf (b)}] The wavelet expansion can be customized to particular applications; 
\item[{\bf (c)}] Because of \textbf{(b)}, the wavelet expansion is more accurate than
the traditional Fourier expansion and also more rapidly converging. 
\end{enumerate}
Wavelet theory has revolutionized the theory of image compression,
the theory of signal processing, and the theory of partial differential
equations. It will be a powerful new tool in the study of epidemiology,
particularly in the analysis of epidemic growth curves.

\section{\textbf{Theoretical Strategy}}

First we propose to build the true wave of an epidemic (through some
harmonic analysis set-up and assumptions) which is otherwise unknown
directly. Then, by assuming a fraction of this constructed wave was
reported out of a true wave, we will then determine how an observed
wave appears. These fractions are variables, so we will have several
patterns of waves representing one true epidemic wave. We will have
to draw conclusions which one of these representations is an ideal
candidate for building a true epidemic. There will be some noise in
our modeling of the epidemic curve, and we will use some noise reduction
techniques before finalizing a pattern (here noise could arise due
to reporting error in disease data).

\begin{figure}
\includegraphics{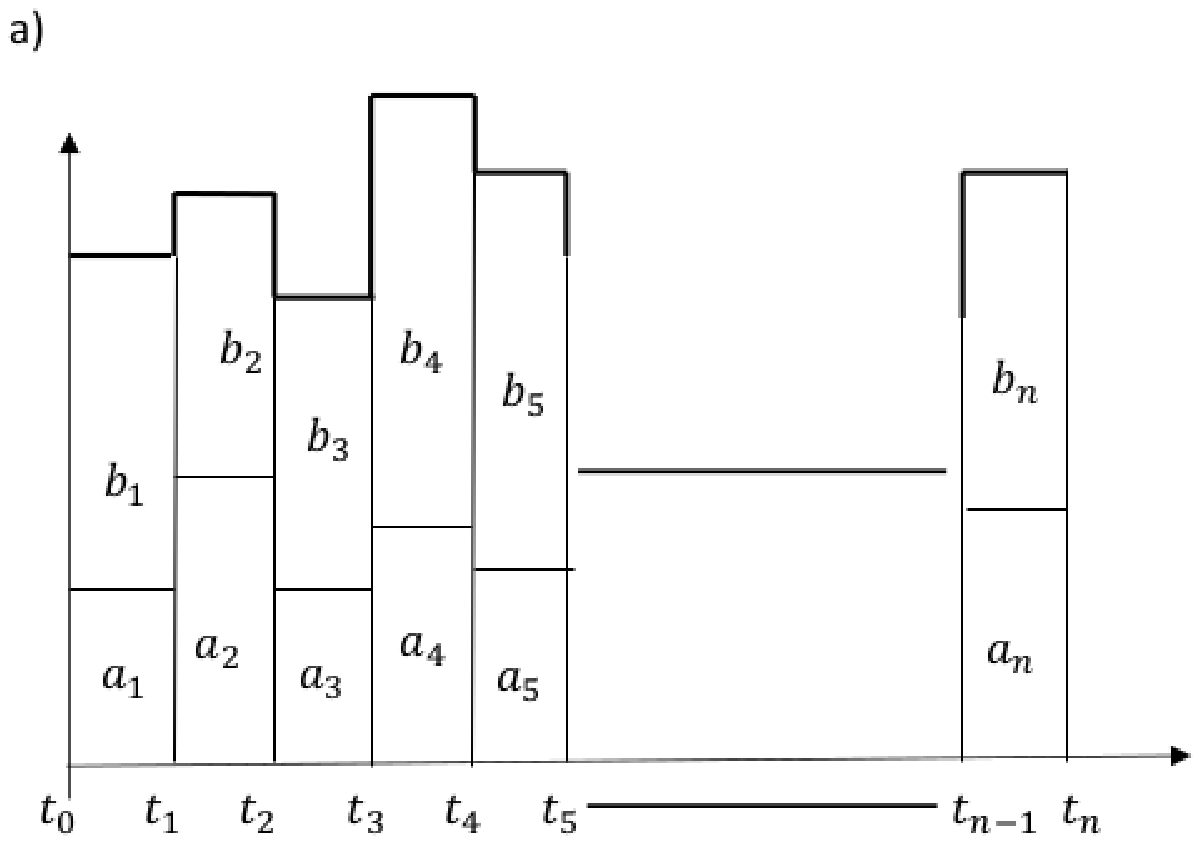}

$ $

\includegraphics{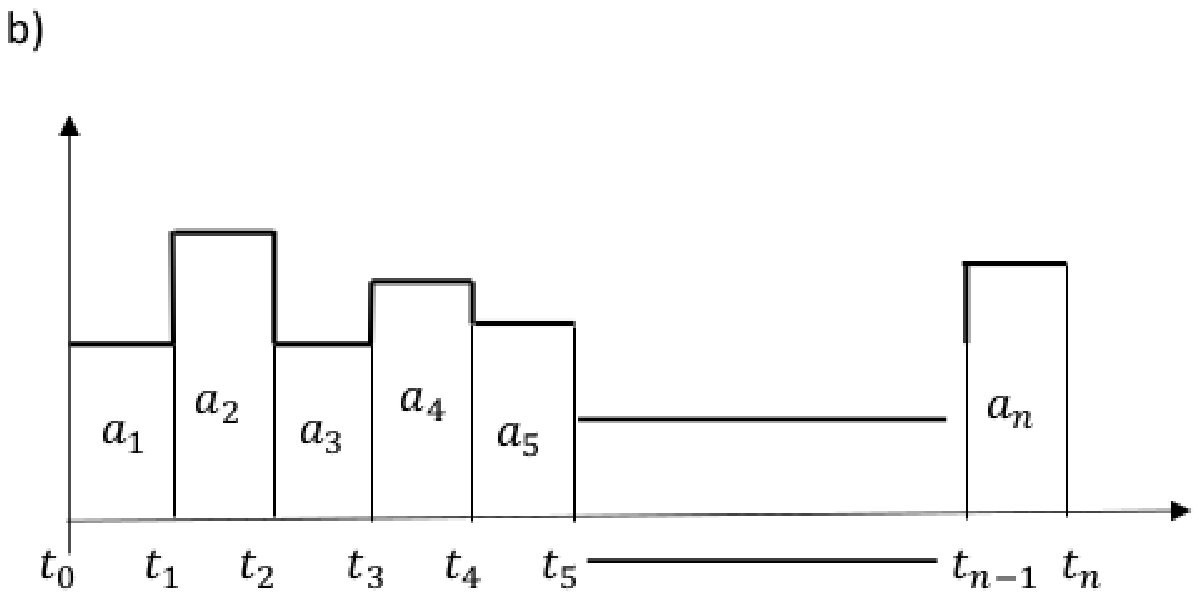}

\caption{\label{Figure1ab}a) True epidemic wave and b) Fractional epidemic
wave. $a_{i}$ values are part of $a_{i}+b_{i}$ values for each $i.$
We have newly introduced the phrase \emph{fractional epidemic waves}
in this work. These fractional epidemic waves concept we are using
with other ideas explained in this paper to develop new ideas related
to \emph{fractional wavelets. }In a sense, fractional wavelets represent
fractions of overall wavelet. This Figure serves as a foundational
concept to link the idea of fractional reporting waves with reporting
errors. }

\end{figure}

\begin{figure}
\includegraphics[scale=0.6]{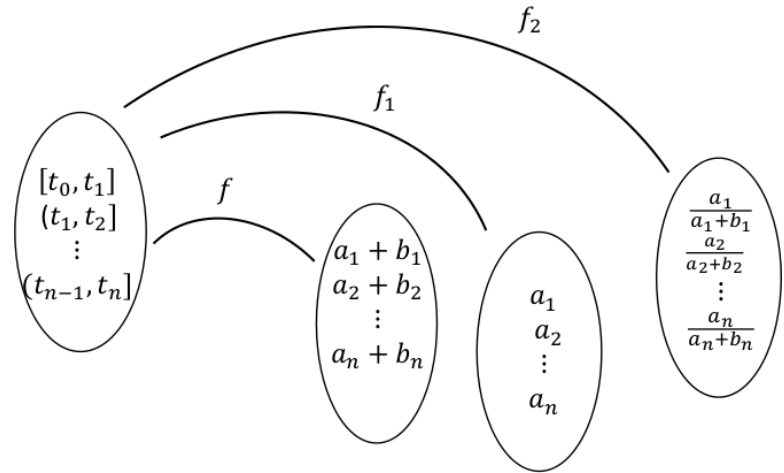}

\caption{\label{Functions of true and fractional}Functions of true and fractional
epidemic waves based on reported and actual time series epidemic data. }

\end{figure}

Suppose an epidemic wave was observed within a time interval $[t_{0},t_{n}]$,
where $t_{n}-t_{0}$ could be in weeks, months, years etc. Suppose
$[t_{0},t_{n}]$ is partitioned into a set $S$ of sub-intervals $\left\{ \left[t_{0},t_{1}\right],\left(t_{1},t_{2}\right],\cdots,\left(t_{n-1},t_{n}\right]\right\} ,$
where $t_{i}-t_{i-1}$ could be in days, weeks depending upon the
situation. Let $a_{i}$ and $b_{i}$ be the number of cases reported
and number of cases those are occurred but not reported, respectively,
within the interval $\left[t_{i-1},t_{i}\right]$ for $i=1,2,\cdots,n.$
Let $f$ be the function whose domain is set of time intervals $\left\{ \left[t_{i-1},t_{i}\right]\mid\forall i\right\} $
and whose range is the set $T=\left\{ a_{i}+b_{i}\mid i=1,2,...,n\right\} $
(See Figure \ref{Figure1ab}). Here $f$ need not be one to one function
because two time intervals within $S$ could have same number of epidemic
cases. Let $f_{1}$ be the function defined as $f_{1}:\left\{ \left[t_{i-1},t_{i}\right]\mid i=1,2,...,n\right\} \rightarrow A,$
where $A=\left\{ a_{i}\mid i=1,2,...,n\right\} $ (See Figure \ref{Functions of true and fractional}).
We call $f_{1}$ as a fractional function of $f.$ The reason we call
this function a fractional function is that it maps each time interval
into corresponding number of reported cases in each of these intervals.
Total fraction of reported cases $\Sigma_{i}a_{i}/\Sigma_{i}(a_{i}+b_{i})\in[0,1]$
during $[t_{0},t_{n}]$ is distributed into $n-$ time intervals.
Whereas,

\[
\sum_{i}\frac{a_{i}}{a_{i}+b_{i}}=\left\{ \begin{array}{c}
n\text{ if all disease cases are reported during \ensuremath{[t_{0},t_{n}]} }\\
<n\text{ if any one interval in \ensuremath{S} there exists under reporting of diseases cases }
\end{array}\right.
\]

Given that the $f_{1}$ is known, the question will be, whether can
we estimate (or speculate) $f$? Once we are able to estimate some
form of $f$, then, how can we test for accuracy of these form(s)
obtained. We could define another fractional function $f_{2}$ as
$f_{2}:\left\{ \left[t_{i-1},t_{i}\right]\mid i=1,2,...,n\right\} \rightarrow\frac{A}{T}$
where $\frac{A}{T}=\left\{ \frac{a_{1}}{a_{i}+b_{i}}\mid i=1,2,...,n\right\} $
and one could attempt (to develop techniques) to estimate (or speculate)
$f$ from $f_{2}.$ In the Figure \ref{Functions of true and fractional},
the fractional epidemic wave pattern is not fully describing true
epidemic wave pattern. Purely from fractional epidemic wave, it is
not easy to speculate true epidemic wave pattern. Additional information
on $b_{i}$ values is needed for better prospects in speculation of
$f.$ 

\begin{figure}
\includegraphics[scale=0.6]{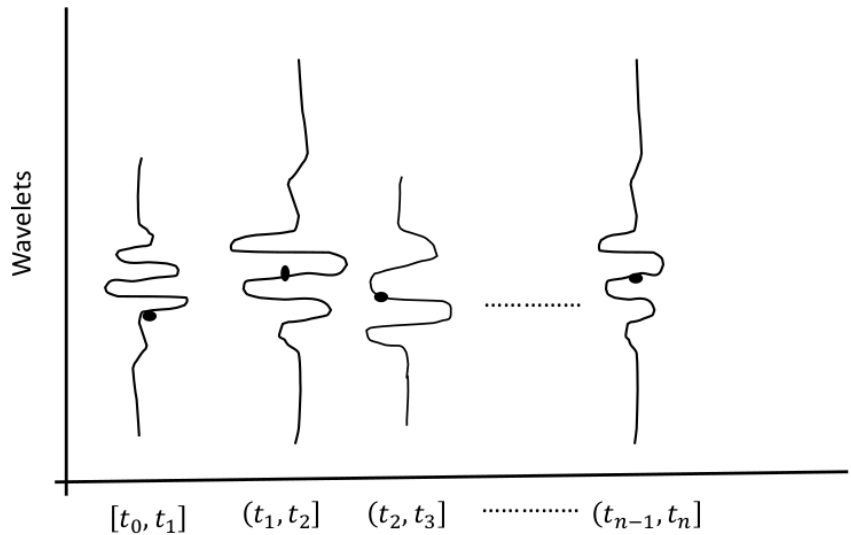}

\caption{\label{fig:Wavelets-constructed-fromsampleddata}Wavelets constructed
from sampled reported data with supports. Black color points on wavelets
represent sampled point (of total reported cases). Each wavelet is
constructed with the pairs of information $\left\{ a_{i},supp(a_{i})\right\} $
available. One of the key technical features is that we are proposing
through this Figure is to construct wavelets within each interval
to quantify the level of reporting cases out of actual cases. }

\end{figure}

\begin{figure}
\includegraphics[scale=0.6]{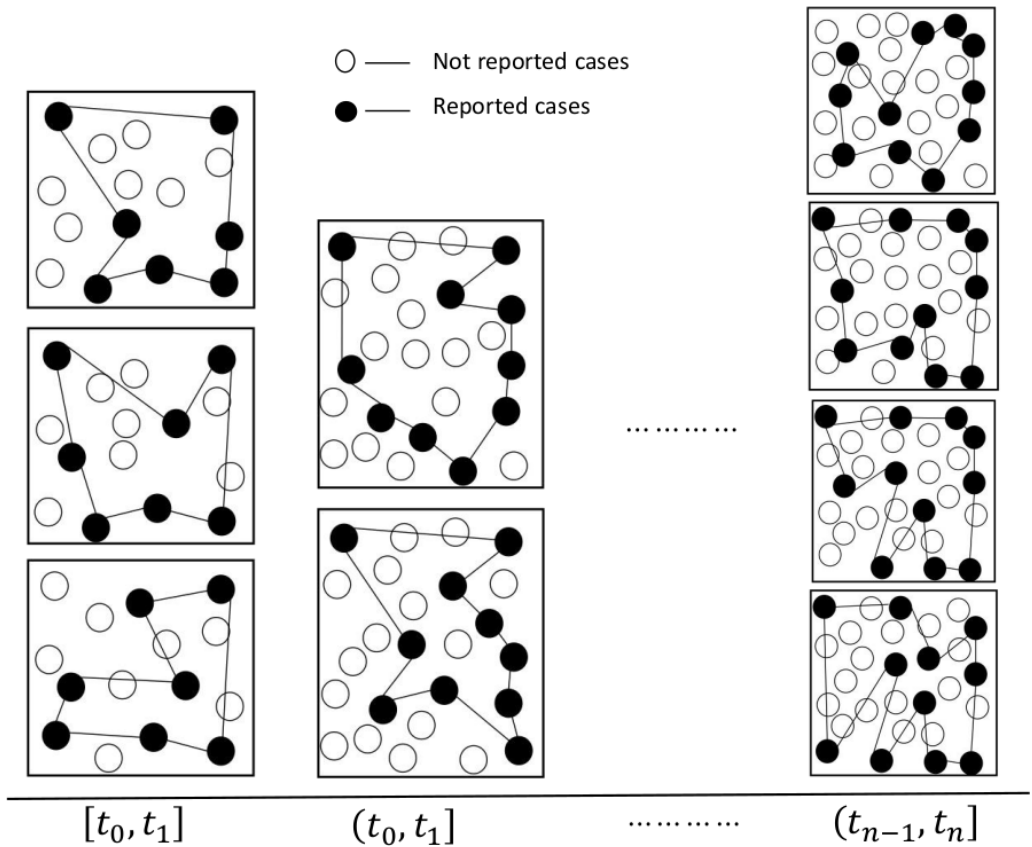}

\caption{\label{fig: sampled cases}Total reported cases within an interval
of time could be formed from a sampled cases out of total cases. We
drew  graphs using black colored filled circles in this Figure to
represent total reported cases in few of the situations out of all
possible diseases reporting patterns. The sample point of reported
cases represents the total reported cases at each time interval. Hence
the size of all  graphs at each interval was kept the same. Similarly,
the size of  graphs between different time intervals are kept different
for demonstration purpose only and actual reported cases between different
time intervals could be constant or not. }

\end{figure}

\begin{figure}
\includegraphics[scale=0.8]{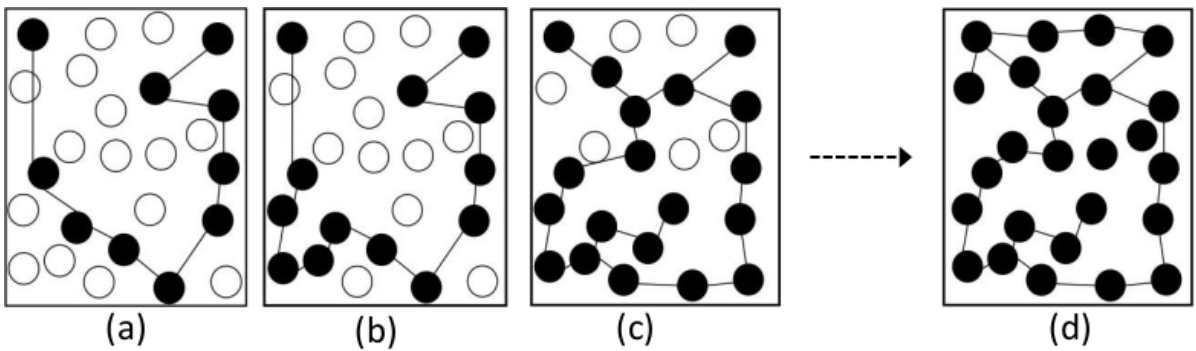}

\caption{\label{fig:Convergence-of-graph}Convergence of graph at sample point
to graph at complete reporting. From sampled point number of reported
cases to the evolution of actual reported cases. This situation arises
due to improved epidemic surveillance.  }

\end{figure}

\begin{figure}
\includegraphics[scale=0.7]{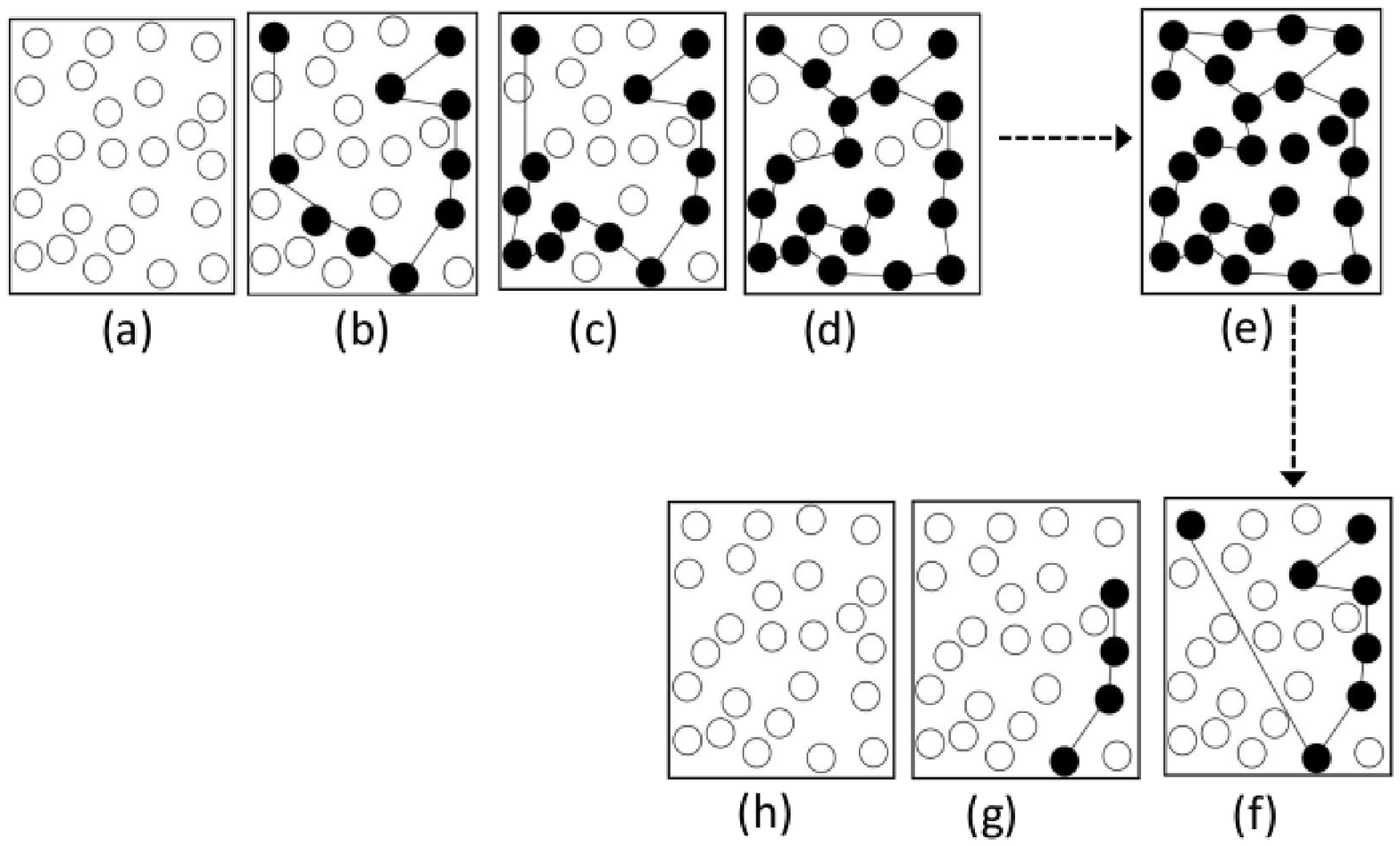}

\caption{\label{waveletgrapgh}Evolution of reporting of epidemic cases and
returning to recovered stage}

\end{figure}

\subsection{Generating wavelets from sampled epidemic data: }

Let us consider availability of data as per the Figure \ref{Figure1ab}
on reported cases. Suppose each point on the y-axis is considered
as a sampled point (out of many sets of plausible reported cases at
that point). We call it a sampled point because we are not sure the
point we obtain at any time intervals $[t_{0},t_{1}]$ and $(t_{i-1},t_{i}]$
for $i=2,3,...,n$ for the reported cases represents a true epidemic
curve and that was one of the main assumptions in this paper. Total
number of reported cases within a time interval is a combination of
cases that were reported to the public health system (See Figure \ref{fig: sampled cases}).
When we know the total number of reported cases within a time interval,
then this number could be resultant of one of the several combinations
of cases reported as shown in the Figure \ref{fig: sampled cases}.
Within each interval the combination of cases reported is unknown
but the total reported cases out of actual diseases cases within each
interval are fixed (because we will take a single point reference
of total reported cases within each interval). These reported cases
are $a_{i}'s$ in Figure \ref{Figure1ab}. Given that there exists
a sampled point within each of the time intervals, and with some support
for each of the $a_{i}$ (say, $supp(a_{i}))$, we will construct
wavelet for each time interval $[t_{0},t_{1}]$ and $(t_{i-1},t_{i}]$
for $i=2,3,...,n$. Sampled point at a time interval we mean, the
final combination of cases reported (out of actual cases) those were
considered as final reporting number for that interval. How we decide
the support is not clear right now, but we will use from a data which
was used to get the sampled point $a_{i}$. Here sampled point does
not necessarily mean statistical sampled point. In the Figure \ref{fig: sampled cases},
for the interval $[t_{0},t_{1}]$, the graph connecting each of the
black circle (vertex) within each square forms a  graph. Although
the sizes of each of these  graphs are same, i.e. $7$, but their
shapes are different and the sampled point is $7.$ The sampled point
cannot be used easily to represent the shape of the  graph unless
location of each node (in this case a physical address or geographical
location of each node) is known. One way to construct the support
could be from the graph associated with each sampled point. Using
the pairs of information $\left\{ a_{i},supp(a_{i})\right\} $ we
will construct wavelets as shown in Figure \ref{fig:Wavelets-constructed-fromsampleddata}.
Sampled points within each interval $[t_{0},t_{1}]$ and $(t_{i-1},t_{i}]$
for $i=2,3,...,n$ gives number of the reported cases and support
constructed from  graphs within each of these intervals. Within each
square or a rectangle in the Figure \ref{fig: sampled cases}, if
the size of a  graph increases to the maximum possible size (i.e.
when all cases are reported), then information to construct corresponding
support increases. 

Let $G_{i}$ be the  graph corresponding to a sampled point $a_{i}$
for the interval $(t_{i-1},t_{i}]$ and suppose $G_{i}^{c}$ be the
 graph with all possible reported cases being reported, then $G_{i}\rightarrow G_{i}^{c}$
($G_{i}$ converges to $G_{i}^{c}$) for all $i$. In $G_{i}^{c}$
the number of vertices are the number of actual disease cases and
edges are connected between closest vertices. In reality, $G_{i}^{c}$
is not possible to draw because we would not be able to observe all
cases. It is challenging to understand before hand what fraction of
the size of $G_{i}^{c}$ would be the size of $G_{i},$and this guess
could give the speed of convergence to $G_{i}^{c}$ from $G_{i}.$
Actual time steps taken from $G_{i}$ to $G_{i}^{c}$ for each $i$
is not constant. We assume there will be a finite number of time steps
to reach from $G_{i}$ to $G_{i}^{c}.$ Usually $c$ is not constant
as well because the error rates vary. So, we let $c_{0}$ corresponds
to complete reporting at $t_{0},$ $c_{1}$ at $t_{1},$ and so on
for $c_{n}$ at $t_{n}.$ Let at $t_{0}$ the graph be $G_{i},$ at
$t_{1}$ the graph be $G_{i}^{t_{1}}$ and so on. The corresponding
sizes of graphs be $\left|E_{i}^{t_{j}}\right|$ for $i=1,2,...,n$
and $j=1,2,...,c_{i},$ and 

\[
\left|E_{i}^{t_{0}}\right|<\left|E_{i}^{t_{1}}\right|<...<\left|E_{i}^{t_{c_{i}}}\right|\:\text{for each }i.
\]

But the inequality,

\[
\left|E_{i}^{t_{j}}\right|<\left|E_{l}^{t_{j+1}}\right|\text{ for some \ensuremath{i\neq l} and }l=1,2,...,n
\]
need not hold. The explanation for these inequalities is as follows:
 graphs within each time interval could converge toward actual disease
cases but the size of the  graph across various time intervals need
not follow any monotonic property because degrees of error in reported
cases could vary over time. Let $G_{i}$ is represented by $(V_{i},E_{i})$
and $G_{i}^{c}$ is represented by $(V_{i}^{c},E_{i}^{c})$ and as
the reporting of diseases cases improves the values of $(V_{i},E_{i})$
increases such that they become exactly $(V_{i}^{c},E_{i}^{c})$ which
we denote here as $G_{i}\rightarrow G_{i}^{c}$. See Figure \ref{fig:Convergence-of-graph}.

We define $\left|E_{i}^{t_{j}}\right|\text{ for }i=1,2,...,n$ as
local steady-state values and $\max_{i}\left(\left|E_{i}^{t_{j}}\right|\right)$
as global steady-state value. 

\begin{figure}
\includegraphics[scale=0.6]{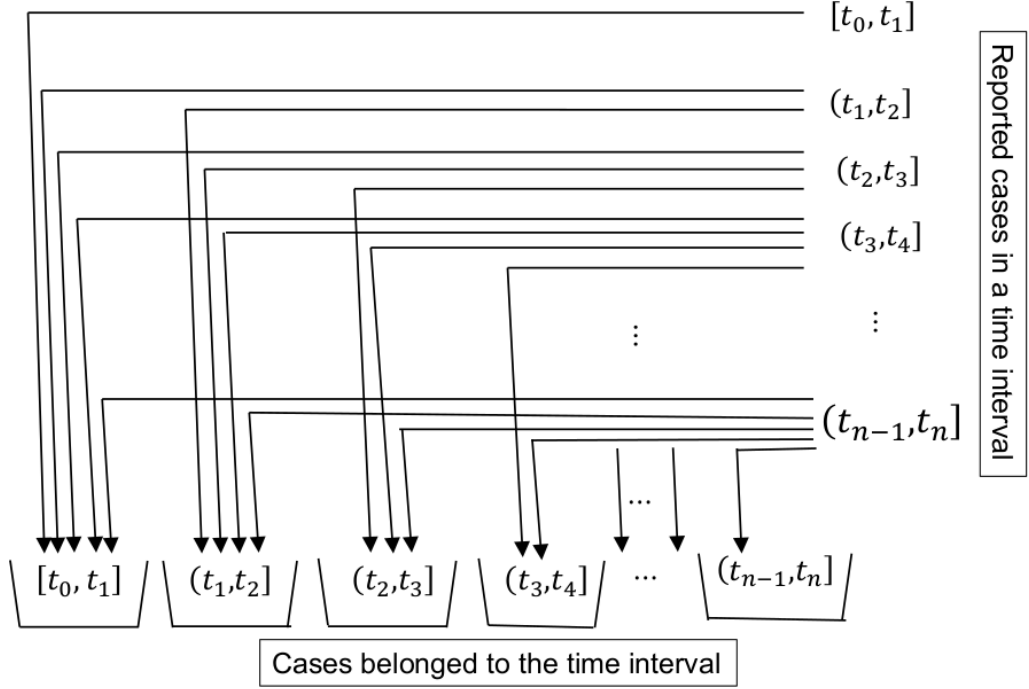}

\caption{\label{fig:Distribution-of-reported}Distribution of reported cases
into present time interval and to past time intervals. Reported cases
found in a time interval in the column are distributed into respective
bins of a time interval as shown through arrows. }
\end{figure}

\begin{figure}
\includegraphics[scale=0.6]{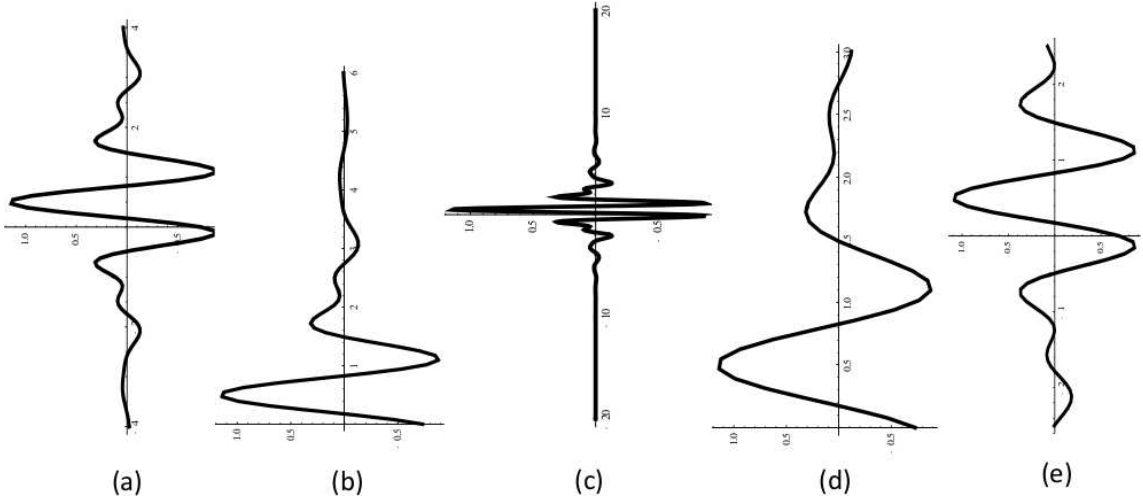}

\caption{\label{fig:Meyer-wavelets}Meyer wavelets of order 3 for various situations
for equally spaced interval of (a) {[}-4,4{]}, (b) {[}0,6{]}, (c)
{[}-20,20,{]}, (d) {[}0,3{]}, (e) with order 10 for {[}-2.5, 2.5{]} }

\end{figure}

\begin{prop}
\label{proposition1}The size of each  graph within $[t_{i-1},t_{i})$
could reach local steady-state and the global steady-state is equal
to the one of the local steady-states. 
\end{prop}

For each $i,$ $\left|E_{i}^{t_{0}}\right|$ and $G_{i}$ are associated
with reported cases. For any $i,$ $\left|E_{i}^{t_{0}}\right|=\left|E_{i}^{t_{c_{i}}}\right|$
then $G_{i}$ and $G_{i}^{t_{c_{i}}}$ are identical, and this situation
refers to complete reporting of disease cases. If $\left|E_{i}^{t_{0}}\right|=\left|E_{i}^{t_{c_{i}}}\right|,$
for any $i$, then local steady-state for this $i$ attained at $t_{0}.$
\begin{rem}
\label{remark2}If $\left|E_{i}^{t_{0}}\right|=\left|E_{i}^{t_{c_{i}}}\right|$
for each $i$ and $\max_{i}\left(\left|E_{i}^{t_{j}}\right|\right)=\left|E_{i}^{t_{0}}\right|,$
then the global steady-state also attains at $t_{0}.$ If $\left|E_{i}^{t_{0}}\right|\neq\left|E_{i}^{t_{c_{i}}}\right|$
for all $i,$ then the global stead-state attains at a time greater
than $t_{0}.$ 
\end{rem}

When $\left|E_{i}^{t_{0}}\right|\neq\left|E_{i}^{t_{c_{i}}}\right|$
for each $i$, then the global steady-state value could provide information
on degree of reporting error to some extent. If $\left|E_{i}^{t_{0}}\right|=\left|E_{i}^{t_{c_{i}}}\right|$
for some $i$, and by chance at this $i,$ the global steady-state
occurs then that wouldn't provide any information on degree of reporting
errors, because at several other $i$ values we will have $E_{i}^{t_{0}}<\left|E_{i}^{t_{c_{i}}}\right|$
and actual total epidemic cases are more than sample epidemic cases. 

Above, statements in the Proposition \ref{proposition1} and in the
Remark \ref{remark2} will alter when multiple reporting exists in
one or more of the time intervals considered. Multiple reporting of
cases is usually defined as reporting of a disease case more than
once and treating it as more than one event of disease occurrence.
When multiple reporting exists at each $i$, then $\max_{i}\left(\left|E_{i}^{t_{j}}\right|\right)\neq\left|E_{i}^{t_{0}}\right|$
is not the global steady-state. A mixed situation where multiple reporting
and under reporting simultaneously exists within the longer time interval
$[t_{0},t_{n}]$ is treated separately. 

With this method, we will develop a series of wavelets. 

Given the information to construct Figure \ref{fig:Convergence-of-graph}(a),
and the rapidity at which this graph evolves from the Figure \ref{fig:Convergence-of-graph}(a)
to the Figure \ref{fig:Convergence-of-graph}(d) (which we might refer
above as \emph{support}) to attain the Figure \ref{fig:Convergence-of-graph}(d)
is known, then combined with the information stored in the Figure
\ref{waveletgrapgh}, we can then construct Figure \ref{fig:Wavelets-constructed-fromsampleddata}.
Within each of the intervals $[t_{0},t_{1}]$ and $(t_{i-1},t_{i}]$
for $i=2,3,...,n$, the information of the Figures \ref{fig:Convergence-of-graph}-\ref{waveletgrapgh},
will be used to construct series of wavelets. 

For example, if some $\Psi(t)$ and some $\Phi(t)$ together describe
the epidemic wave of a true epidemic, and if pairs of functions$\{(\Psi_{i}(t),\Phi_{i}(t))\}$
for $i=1,2,...,n$ represent the epidemic wave of those representing
fractions of this true epidemic, then one of our central ideas is
in determining which of these fractional wave is closest to the true
epidemic. Usually data/information to construct a couple of such fractional
wavelets could be observed in an emerging epidemic, say $(\Psi_{a}(t),\Phi_{a}(t))\}$
and $(\Psi_{b}(t),\Phi_{b}(t))$, so the first step is to construct
these pairs of wavelets. These fractional wavelets are constructed
on partial data (partial in the sense that observed data on disease
cases in an emerging epidemic is not complete). The question we are
attempting is: can we predict $(\Psi(t),\Phi(t))$ from either one
of or from both of the fractional wavelets. {[}\textbf{Note:} There
is no terminology of ``fractional wavelet'' in the literature, but
we are calling $(\Psi_{a}(t),\Phi_{a}(t))$ and $(\Psi_{b}(t),\Phi_{b}(t))$
the fractional wavelet{]}. For this, let us consider Meyer wavelets
which are readily available and could be a good first step to start
with to explain our epidemic situation. We will define Meyer wavelet
and briefly describe them below:

The Meyer wavelet is an orthogonal wavelet created by Yves Meyer.
It is a continuous wavelet, and has been applied to the study of adaptive
filters, random fields, and multi-fault classification. 
\begin{defn}
The Meyer wavelet is an infinitely differentiable function that is
defined in the frequency domain in terms of a function $\nu$ as follows:

\[
\Psi(\omega)=\begin{cases}
\begin{array}{c}
\frac{1}{\sqrt{2\pi}}\sin\left(\frac{\pi}{2}\nu\left(\frac{3\left|\omega\right|}{2\pi}-1\right)\right)e^{j\omega/2}\\
\frac{1}{\sqrt{2\pi}}\cos\left(\frac{\pi}{2}\nu\left(\frac{3\left|\omega\right|}{2\pi}-1\right)\right)e^{j\omega/2}\\
0
\end{array} & \begin{array}{cc}
\begin{array}{c}
\text{if}\\
\text{if}\\
\text{if}
\end{array} & \begin{array}{c}
2\pi/3<\left|\omega\right|<4\pi/3\\
4\pi/3<\left|\omega\right|<8\pi/3\\
\text{otherwise.}
\end{array}\end{array}\end{cases}
\]

Here

\[
\nu(x)=\begin{cases}
\begin{array}{c}
0\\
x\\
1
\end{array} & \begin{array}{cc}
\begin{array}{c}
\text{if}\\
\text{if}\\
\text{if}
\end{array} & \begin{array}{c}
x<0\\
0<x<1\\
x>1.
\end{array}\end{array}\end{cases}
\]

There are other possible choices for $\nu.$ 

The Meyer scaling function is given by 

\[
\Phi(\omega)=\begin{cases}
\begin{array}{c}
\frac{1}{\sqrt{2\pi}}\\
\frac{1}{\sqrt{2\pi}}\cos\left(\frac{\pi}{2}\nu\left(\frac{3\left|\omega\right|}{2\pi}-1\right)\right)\\
0
\end{array} & \begin{array}{cc}
\begin{array}{c}
\text{if}\\
\text{if}\\
\text{if}
\end{array} & \begin{array}{c}
\left|\omega\right|<2\pi/3\\
2\pi/3<\left|\omega\right|<4\pi/3\\
\text{otherwise.}
\end{array}\end{array}\end{cases}
\]

Of course it holds, as usual, that

\[
\underset{k}{\Sigma}\left|\hat{\Phi}\left(\omega+2\pi k\right)\right|^{2}=\frac{1}{2\pi}
\]

and

\[
\hat{\Phi}(\omega)=m_{0}(\omega/2).\hat{\Phi}(\omega/2)
\]

for some $2\pi-$periodic $m_{0}(\omega/2).$ Finally, 

\[
\begin{array}{ccc}
\Psi(\omega) & = & e^{i\omega/2}\overline{m_{0}(\omega/2+\pi)}\hat{\Phi}(\omega/2)\;\qquad\quad\quad\;\;\\
 & = & e^{i\omega/2}\underset{k}{\Sigma}\overline{\hat{\Phi}\left(\omega+2\pi(2k+1)\right)}\hat{\Phi}(\omega/2)\quad\quad\;\\
 & = & e^{i\omega/2}\left(\hat{\Phi}\left(\omega+2\pi\right)+\hat{\Phi}\left(\omega-2\pi\right)\right)\hat{\Phi}(\omega/2).
\end{array}
\]

It turns out that the wavelets

\[
\Psi_{j,k}(x)=2^{j/2}\Psi(2^{j}x-k)
\]

form an othonormal basis for the square integrable functions on the
real line. 
\end{defn}

One proposition that could be formed is ``if a wavelet is constructed
on the partial data of a particular series of events in a population,
then this wavelet will not be fully compared with a wavelet constructed
from the full data series of all events in the same population.''
Building a measure associated with these two wavelets�is interesting
and there could be several such measures based on the level of completeness
in the data. Because we are dealing with true versus reported disease
cases this measure (a set of points each representing a distance between
true and observed cases) could be termed the \textit{error} in reporting
of disease cases. These kind of measures will be very helpful (such
measures after further filtration can be useful for practical epidemiologists).
Instead of constructing wavelets for the overall epidemic duration,
we will construct wavelets within intervals $[t_{0},t_{1}]$ and $(t_{i-1},t_{i}]$
for $i=2,3,...,n$ as described in the Figure \ref{fig:Wavelets-constructed-fromsampleddata}.
As the reported cases within an interval improve, described as in
the Figure \ref{fig:Convergence-of-graph}, wavelets configuration
improves. Each of the fractional wavelets obtained from partial data
will be updated using the information shown in the Figure \ref{fig:Distribution-of-reported}.
Meyer wavelets for various equally spaced intervals are demonstrated
in the Figure \ref{fig:Meyer-wavelets}. 

\subsubsection{Computation}

Suppose a sample point is obtained for an interval $(t_{i-1},t_{i}]$.
An improvement of reported cases is tried to ascertain from the data
obtained in the subsequent time intervals. One way to update this
is from future epidemic cases those were infected and or diagnosed
for the period $(t_{i-1},t_{i}]$ but made available during any of
the intervals $(t_{i},t_{i+1}]$ for $i=2,3,...,n-1.$ That is, sum
of the epidemic cases those were reported during $(t_{i-1},t_{i}]$
and those reported during each of the future time intervals $(t_{i},t_{i+1}]$
for $i=2,3,...,n-1$ and belong to the interval $(t_{i-1},t_{i}]$
will be treated as improved number of reported cases for the interval
$(t_{i-1},t_{i}]$. We will update reported number of cases in a previous
interval from future available reported cases that was associated
with previous time interval. Hence, the evolution of the data for
the interval $(t_{i-1},t_{i}]$ can be used to construct graphs shown
in Figure \ref{fig:Convergence-of-graph}. Since this evolution is
assumed to observe for a long period, it is assumed that $G_{i}$
will be convergent to $G_{i}^{c}$ approximately. As an epidemic progresses,
we update the intervals $(t_{i-1},t_{i}]$ with newly available information
during $(t_{i},t_{i+1}]$ and as $i$ approaches $n$ then those intervals
nearby to $n$ will have less chance of evolution or less chance of
up-gradation (due to truncation effect). Once the reporting numbers
are complete, we will similarly study recover stage of an epidemic
and hence collect the data to compute the Figure \ref{waveletgrapgh}.
Accumulation of old cases in new intervals are distributed back to
respective time intervals is schematically described in the Figure
\ref{fig:Distribution-of-reported}. This procedure will update the
reported cases in past as long as at least one reported case is observed
in the present time interval that belongs to one of the past time
interval. Based on the location of the reported cases the graphs constructed
will be updated as well. Hence for each present time interval we observe
a reported case that belongs to one of the past time interval, the
fractional wavelets will become graphically closer to the complete
(or true) wavelet for that time interval. 

One way to assess closest fractional wavelet is to compare some features
of fractional wavelets with a PDE/ODE model of an emerging epidemic. 

\section{\textbf{What we can achieve through such an analysis?}}

We provide a group of epidemic growth scenarios inspired from the
harmonic analysis set-up. A couple of scenarios from this could represent
true epidemic growth curves. {[}We will have to evolve a strategy
to short list a couple of plausible true scenarios.{]} With this,
we will be in a position to assess the level of under-reporting in
a particular epidemic. So the strategy we are proposing could be beneficial
in not only building a true epidemic but also assessing the level
of reporting error in an epidemic. We provide the true epidemic with
noise to illustrate our claim.

In addition to the gain in construction of true epidemic, we also
propose new methods that blend harmonic analysis with dynamical systems
and sampling strategy. In this way, harmonic analysis is used to bridge
the gap between unknown and known information in disease epidemiology.
Suppose we determine one of the fractional wavelets, say $(\Psi_{a}(t),\Phi_{a}(t))$,
is closest to the true wavelet; then finding a measure which is the
difference between $(\Psi_{a}(t),\Phi_{a}(t))$ and $(\Psi(t),\Phi(t))$
will complete mapping of the epidemic at time $t.$ However determining
which one of the fractions $(\Psi_{i}(t),\Phi_{i}(t))$ is closest
to the true is not so easy. But if there are no significant multiple
reporting of disease cases then the largest fractional wavelet could
be assumed to be the one with shortest measure {[}Note: Still we need
to provide more clarity on strategy to determine the closest available
fractional wavelet{]}.

We are trying to use wavelets to extract full data from a partial
data. We have argued how various combinations of partial data can
be used for discrete constructions which in turn form a supporting
information to construct wavelets. These two aspects makes our proposed
work very innovative. In summary, what we are trying to develop through
this paper is, given that we have partial data of an event (here we
mean event of reporting of disease cases), we will construct the complete
event data. Wavelets, in this work, are occupying key role in processing
of built-up or accumulated data to build complete event data. The
event here is the reported number of cases in a time interval and
these reported cases represent only partial number of actual epidemic
cases.Through this paper we demonstrated a method of improving of
partial data to close to a data which could be complete. How do we
plan to update our data reported in an interval and bring it closer
to actual number of disease cases is described in this paper. 

This exploration will assist in better visualization of any emerging
epidemic spread in a more realistic sense. We also believe that our
methods could provide additional tools for those epidemic modelers
who frequently use modeling tools such as ODE and PDE to begin with.
We are not only looking for academic development through this project,
but, also a clear non-trivial body of techniques for applications
of harmonic analysis that has never been seen before in-terms of developing
epidemic analysis. We plan to come up with some interesting insights
on how to construct wavelets for medical applications. The bottom
line is that we will be able to help public health planners for better
management and courses of action during emerging epidemics. The kind
of analysis we present here to fill missing pieces of epidemic reporting
information can be applied to other areas, for example, constructing
total rhythm of a heart beat from partial information, etc 

\section{\textbf{Questions still remain }}

Within what span of time we can generate a true epidemic from its
emergence using the harmonic analysis set-up? Can we predict the full
picture of an epidemic from only partial data? Can we measure the
validity and the accuracy of an epidemic growth curve? Can we measure
the timeliness of our analysis?

\section*{\textbf{Global View}}

We have identified a gap in the methods of understanding true epidemic
growth and spread and tried to address this by proposing a novel method.
We are proposing through this study that wavelets could offer a road
map closer to finding a practical solution (we are aware that a perfect
solution is impossible by any method because some of the disease cases
in any situation are never reported). Technical aspects of the story
line was depended on construction of discrete graphs and \emph{fractional
wavelets}. Fractional wavelets are newly introduced to the literature
through this study. A solution through wavelets is also not trivial
because there is no ready-made set of wavelets available which will
offer a timely road-map. So we have introduced a novel strategy. Hence
we argue that our approach will help to come closer to our aims of
understanding epidemics in a more accurate and timely fashion. As
a bi-product, we can develop techniques for data scientists to analyze
disease surveillance.

\end{document}